\documentclass[twocolumn,aps,prl,amsmath,epsfig,superscriptaddress,showpacs,floatfix] {revtex4}

\usepackage{amssymb}

\usepackage[dvips]{graphicx}

\begin{document}
%\draft

\newcommand{\lxpc} {Li$_{x}$ZnPc }
\newcommand{\lp} {Li$_{1.85}$ZnPc }
\newcommand{\lppc} {Li$_{3.75}$ZnPc }
\newcommand{\lpc} {Li$_{0.5}$MnPc }
\newcommand{\lmpc} {Li$_{x}$MnPc }
\newcommand{\etal} {{\it et al.} }
\newcommand{\ie} {{\it i.e.} }
\newcommand{\ip}{${\cal A}^2$ }

\hyphenation{a-long}

\title{Strong electronic correlations in Li$_x$ZnPc organic metals}

\author{M. Filibian}
%$\footnote{e-mail:
%filibian@fisicavolta.unipv.it}$}
\author{P. Carretta}
\author{M. C. Mozzati}

\affiliation{Department of Physics \lq\lq A. Volta\rq\rq,
University of Pavia, Via Bassi 6, I-27100 Pavia, Italy}

\author{P. Ghigna}
\affiliation{Department of Physical Chemistry  "M. Rolla",
University of Pavia, V.le Taramelli 16, I-27100 Pavia, Italy}

\author{G. Zoppellaro}
\author{M. Ruben}
\affiliation{Institut f$\ddot{u}$r Nanotechnologie,
Forschungszentrum Karlsruhe, 76021 Karlsruhe, Germany}

\date{\today}
%\maketitle
\widetext
%%%%%%%%%%%%%%%%%%%%%%%%%%%%%%%%%%%%%%%%%%%%%%%%%%%%%%%%%%%%%%%%%%%%%%%%%%%%%%%%

      \begin{abstract}

Nuclear magnetic resonance, electron paramagnetic resonance and
magnetization measurements show that bulk Li$_x$ZnPc are strongly
correlated one-dimensional metals. The temperature dependence of
the nuclear spin-lattice relaxation rate $1/T_1$ and of the static
uniform susceptibility $\chi_S$ on approaching room temperature
are characteristic of a Fermi liquid. Moreover, while for $x\simeq
2$ the electrons are delocalized down to low temperature, for
$x\rightarrow 4$ a tendency towards localization is noticed upon
cooling, yielding an increase both  in $1/T_1$ and $\chi_s$. The
$x$-dependence of the effective density of  states at the Fermi
level $D(E_F)$ displays a sharp enhancement for $x\simeq 2$, at
the half filling of the ZnPc lowest unoccupied molecular orbitals.
This suggests that Li$_x$ZnPc is on the edge of a metal-insulator
transition where enhanced superconducting fluctuations could
develop.

\end{abstract}
%%%%%%%%%%%%%%%%%%%%%%%%%%%%%%%%%%%%%%%%%%%%%%%%%%%%%%%%%%%%%%%%%%%%%%%%%%%

 \maketitle

Organic molecular crystals have been the subject of an intensive
research activity for decades in view of their complex electronic
properties and several technological applicabilities \cite{intro}.
In these crystals the bandwidth W is narrow (0.1$\div$1 eV) and
they are usually insulators, as charge delocalization can be
achieved only when W is larger than the on-site Coulomb repulsion
U. Nevertheless, when U/W$\simeq 1$ and the local energy scales
become relevant, it is still possible to observe unexpected
phenomena as the enhancement of superconducting fluctuations. In
this respect alkali-doped fullerenes A$_x$C$_{60}$ are
paradigmatic. For $x=3$ the $\pi$ Lowest Unoccupied Molecular
Orbital (LUMO) band is half-filled and, despite of the significant
correlations, they are metals featuring a superconducting
transition at a surprisingly high critical temperature
\cite{Gunn}. The unconventional superconductivity in A$_x$C$_{60}$
has been ascribed to the competition between the Jahn-Teller (JT)
coupling of the electrons with the intramolecular phonon modes
 \cite{Tosattiscience} and the local exchange J (Hund coupling),
representing the relevant energy scales when U/W$\simeq$1. In this
limit the system behaves as if it had a reduced bandwidth $W^*$
and, as far as the effective exchange $J_{eff}=J-E_{JT}/4$ remains
of the order of $W^*$, even in
presence of a high Coulomb repulsion an attractive electron pairing can arise \cite{Tosattiscience}.\\
Recently it has been pointed out that also the electronic
properties of the insulating transition metal phthalocyanines (MPc
where Pc=C$_{32}$H$_{16}$N$_8$) thin films can be affected by
alkali ions intercalation \cite{Morpurgo}. MPc molecules consist
in a planar organic macrocycle hosting a metal ion in the central
cavity and form typical slipped-stacked chains. Along the
stacking direction $b$ the Pc antibonding $\pi^*$ LUMOs have the strongest overlap and
form a nearly one-dimensional (1D) narrow band \cite{L}. The results
reported in Ref. \cite{Morpurgo} indicate that A$_x$MPcs become
metallic when the doubly degenerate 2$e_g$ LUMO is half-filled for
$x\simeq 2$ and become insulating again when the band is completely
filled for $x=4$. The possibility that superconductivity could
develop in A$_x$MPc was suggested by Tosatti \etal \cite{Tosatti}.
In fact, they pointed out that A$_x$MPc should be at the edge of a
Metal to Insulator Transition (MIT) \cite{Tosattiscience}. Then
U/W$\sim 1$ and the 2$e_g$ electrons could strongly couple with
the intramolecular $B$ phonon modes \cite{ramangenerale} to yield
$J_{eff}\sim W^*$, as in fullerides. Nevertheless, recent Photo
Electron Spectroscopy results \cite{Giovann, betti}
seem to indicate that K-doped MPc thin films are insulating,
questioning the results reported in Ref. \cite{Morpurgo}.

In this work we present an experimental study of the electronic
properties of bulk $\beta$-Li$_x$ZnPc by means of Nuclear Magnetic
Resonance (NMR), Electron Paramagnetic Resonance (EPR) and Squid
Magnetometry. The synthesis of $\beta$-A$_x$MPc powder samples was
attained recently by means of liquid-phase methods \cite{Yasu}.
These compounds form molecular crystals in which the cell
parameters of the pure $\beta$ form are modified in order to
accommodate the alkali ions in intrastack positions, bound to the
pyrrole bridging nitrogens of adjacent Pc rings \cite{Yasu}. The
$\beta$ form is preferred by MPc in the bulk state to the $\alpha$
form typical of thin films. It must be remarked that the liquid
phase synthesis is a low energy and slow process, yielding a
better control of homogeneity and defects than the higher energy
ionic implantation processes adopted to dope thin films. Thus, the
properties of the powder samples produced by means of the
liquid-phase process used in this work are definitely
reproducible.\\
\lxpc powders were synthesized under inert atmosphere, following
the procedure published by Taguchi \etal in Ref.~\onlinecite{Yasu}
with a few modifications. ZnPc powders were freshly synthesized,
repeatedly filtered and purified by sequential vacuum sublimation.
The blueish residue of Li$_x$ZnPc, obtained by dissolving ZnPc in 1.6 M
n-BuLi in hexane of the appropriate molarity, was dried and it
underwent annealing procedure into NMR tubes with heating at 240
$^{\circ}$C under vacuum (6.8 $\times$ 10$^{-2}$ mbar) for 3-4
days. This heating procedure ensures formation of the $\beta$
polymorph. The NMR tubes were then sealed to avoid oxidation. A
small fraction (a few percent) of LiPc was found to be present in
some samples. The air sensitivity and the powder morphology
prevents one from performing standard transport measurements.

NMR measurements were performed by using standard radiofrequency
(RF) pulse sequences.  The lithium content $x$ was determined from
the relative intensity of $^7$Li and $^1$H NMR signals, namely
from the echo intensity $E(2\tau)$ extrapolated to a delay
$\tau\rightarrow 0$. The values of $x$ derived were consistent
with the doping estimated by Induced Coupled Plasma analysis.
$^7$Li NMR spectra were obtained from the Fourier transform of
half of the echo signal after a $\pi/2-\tau -\pi/2$ pulse
sequence. They showed the typical features of a I=3/2 powder
spectrum (see the inset to Fig. 1) with a quadrupole coupling
frequency $\Delta \nu_Q \simeq$ 44 kHz. The quadrupole frequency
and the lineshape are temperature (T) independent up to room
temperature (RT), namely the local electric-field gradient does
not change upon varying T. This is a neat indication that lithium
does not diffuse and that it is rather strongly bound to the Pc
molecules.

Nuclear spin-lattice relaxation rate $1/T_1$ was estimated from
the recovery of nuclear magnetization $m(\tau)$ after a saturating
RF pulse sequence. $^7$Li recovery law was observed to be a
stretched exponential, namely $y(\tau)\equiv 1-
m(\tau)/m(\tau\rightarrow \infty)= exp(-(\tau/T_1)^{\beta})$, with
$\beta\simeq 0.65$ over the entire T range. The stretched
character of the recovery laws indicates a distribution of
relaxation rates with an average relaxation $1/T_1$ \cite{Tedoldi}.
This likely originates from the anisotropic hyperfine coupling
tensors or from some disorder in Li distribution.
%%%%%%%%%%%%%%%%%%%%%%%%%%figure 5 E 6
\begin{figure}[b]
  \begin{center}
  \vspace {0cm}
  \includegraphics*[angle=270,scale=0.31]{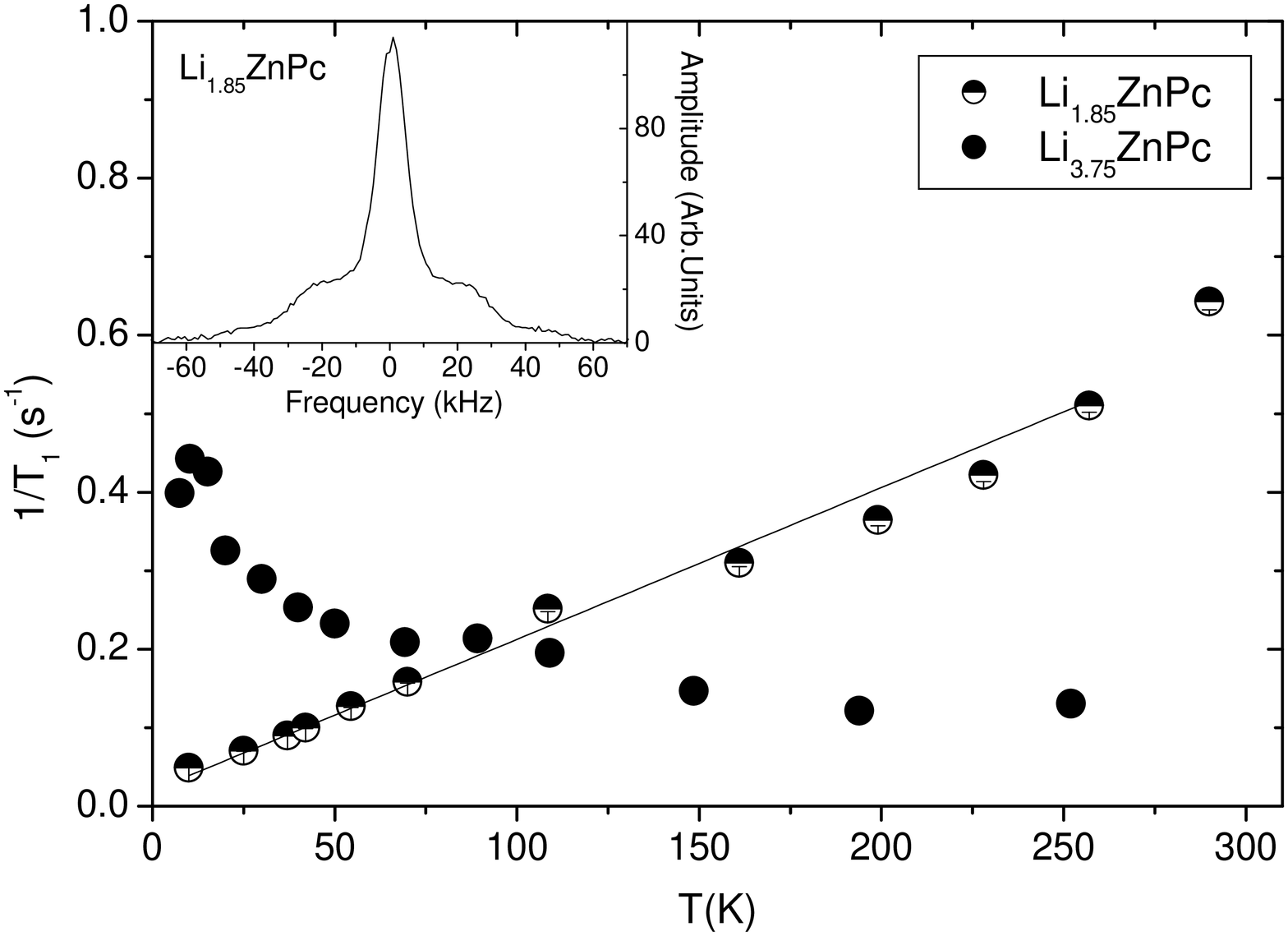}
  \caption{$^{7}$Li $1/T_1$ in
\lxpc as a function of T. The solid line is the
  fit according to Eq.(\ref{Korringaznpc}). In the inset $^{7}$Li NMR powder spectrum in Li$_{1.85}$ZnPc, for $T= 293$ K
  is reported.}\label{1overt1}
  \end{center}
  \end{figure}
%%%%%%%%%%%%%%%%%%%%%%%%%%%%%%%%%%%%
$^7$Li $1/T_1$ in \lxpc powders was found to be about two orders of magnitude smaller than in Li$_x$MnPc, which is insulating \cite{mio}.
Since $^7$Li hyperfine coupling A in \lxpc ($\simeq 2$ kGauss, see later on) is much larger than that in \lmpc
($\sim$100 Gauss), the significant decrease of $^7$Li $1/T_1$ in \lxpc must reflect an increase in the frequencies
of the spin fluctuations by several orders of magnitude, as one would expect when the electrons become
delocalized. The T dependence of $^7$Li $1/T_1$ relaxation rates in \lp and \lppc powder samples is shown in
Fig.\ref{1overt1}.
%%%%%%%%%%%%%%%%%%%%%FIG3%%%%%%%%%%%%%%%%%%%%%%%%%%%%%%%%%%%%%%%%%%%%
\begin{figure}[b]
 \begin{center}\vspace {0cm}
  \includegraphics*[angle=270,scale=0.32]{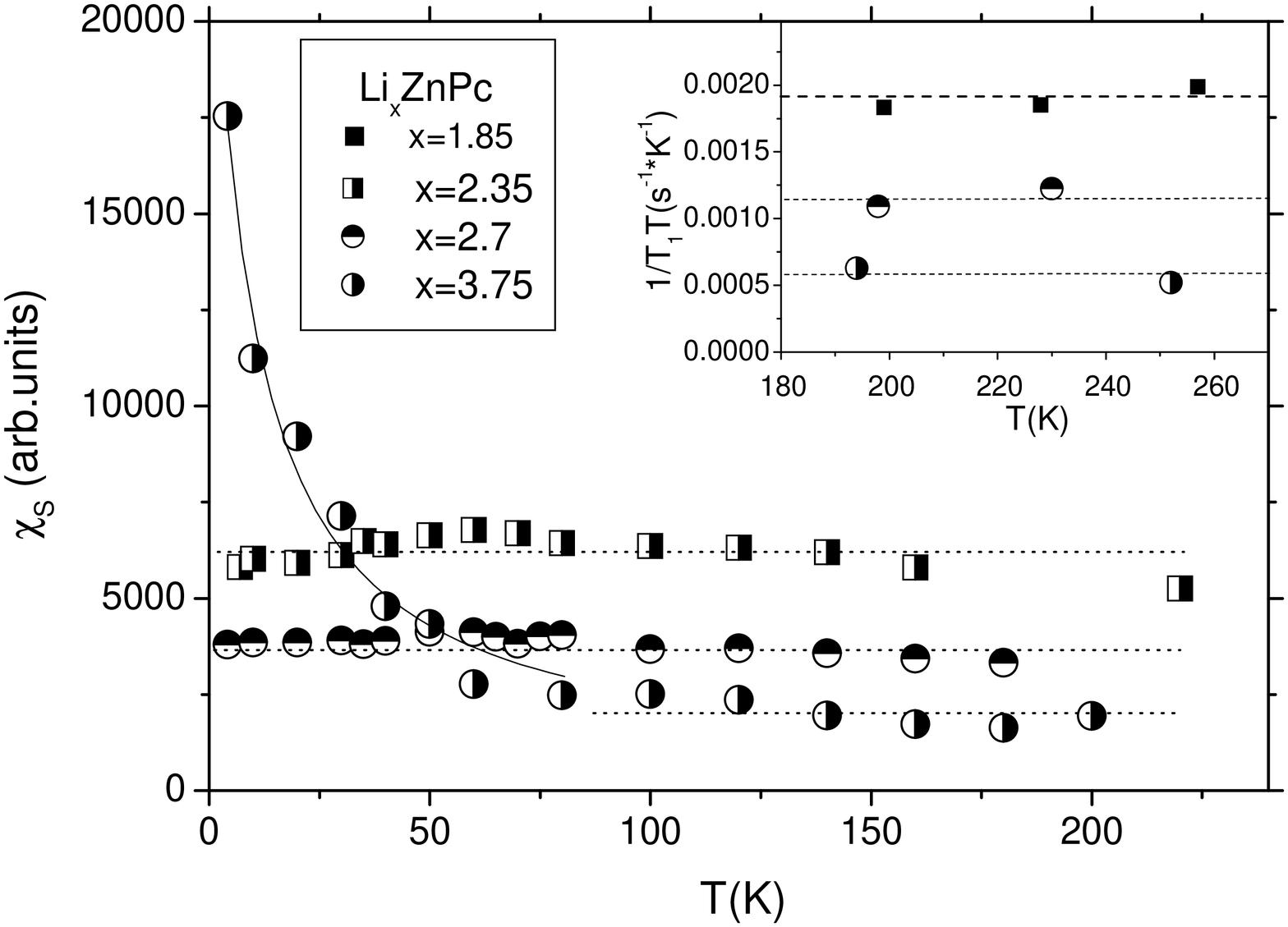} \caption{ Spin susceptibility, derived from the
integral of the EPR spectra, in \lxpc compounds as a function of
T. The dotted lines are guides to the eye marking the high T limit
of the susceptibility. The solid line is a Curie-Weiss fit. The
inset shows the $^{7}$Li $1/T_1T$ values at high T marked by the
dotted lines.}\label{epr}
\end{center}
\end{figure}
%%%%%%%%%%%%%%%%%%%%%%%%%%%%%%%%%%%%%%%%%%%%%%%%%%%%%%%%%%%%%%%%%%%%%%
$1/T_1(T)$ shows significant changes upon varying the lithium content owing to the evolution of the electronic
correlations. In Li$_{1.85}$ZnPc  $^{7}$Li $1/T_1(T)$ is linear down to the lowest T, as expected for a Fermi
Liquid \cite{Slichter}
\begin{equation}
\frac{1}{T_1}=  \frac{\gamma_N^2}{2} A^2 k_B\hbar D(E_F)^2 T=C_0
\chi_S^2 T ,\label{Korringaznpc}
\end{equation}
where $D(E_F)$ is the density
of states at the Fermi level and $\chi_S$ the static uniform spin
susceptibility. On the contrary, in Li$_{3.75}$ZnPc  $1/T_1$
deviates from Eq. (\ref{Korringaznpc}) on decreasing T and
rather diverges at T$\simeq$10 K. The rise in $1/T_1$ below T=150
K recalls the one observed in other 1D strongly
correlated metals as the Bechgaard salts \cite{Jerome}. In a
quasi-1D metal, owing to the Fermi surface topology, the
low-energy excitations contributing to nuclear spin-lattice
relaxation are essentially the ones at $\vec q\simeq 0$ and at
$\vec q\simeq 2\vec k_F$, then one can write \cite{nmrbechgard}
    \begin{equation}
    \frac{1}{T_1}(T)= C_0 \chi_S(T) ^2 T +
    C_1(2\vec{k}_F)T\chi(2\vec{k}_F,\omega_L).\label{beechcomplt1}
    \end{equation}
In Eq. (\ref{beechcomplt1}) the first term corresponds to the
Eq. (\ref{Korringaznpc}), where the static uniform susceptibility $\chi_S(T)$
is in general T-dependent due to the strong electronic
correlations, whereas in a weakly correlated metal $\chi_S(T)$
should be Pauli-like, i.e. nearly constant. Hereafter we will show
that, even if in \lxpc $U/W\rightarrow 1$, as far as the electrons
remain delocalized $\chi_S(T)$ is almost T-independent.

$\chi_S(T)$ for \lxpc powders was estimated from the area of the
EPR spectra, recorded with an X-band spectrometer. The spectra
were characterized by an asymmetric lineshape with a broadening of
a few hundreds Gauss. Further details on the EPR spectra will be
reported elsewhere. Figure \ref{epr} shows that $\chi_S$ for
$x\simeq2$ is nearly constant over all the explored T range,
namely it is Pauli-like and arises from the delocalized electrons
filling the LUMO band, in agreement with what was found from NMR
$1/T_1$ measurements. Moreover, it is noticed that upon increasing
$x$ $\chi_S$ decreases, exactly as $1/T_1T$ in the high T limit
(see the inset to Fig. 2). Eventually, for $x\simeq 3.75$,
$\chi_S$ grows on cooling below T$\sim 150$ K. This rise can be
interpreted either as a consequence of electron localization or
can be simply due to a few percent of impurities \cite{forro}.
Nevertheless, the precise origin of this localization is still
uncertain. Magnetization (M) measurements performed in
Li$_{1.85}$ZnPc using an MPMS-XL7 Quantum Design SQUID
magnetometer are consistent with the EPR measurements. The
macroscopic susceptibility, defined as $\chi= M/H$ with H the
intensity of the applied magnetic field (inset of Fig. 3), is
practically constant, except for a low-T Curie tail ascribable to
impurities.
%%%%%%%%%%%%%%%%%%%%%FIG4%%%%%%%%%%%%%%%%%%%%%%%%%%%%%%%%%%%%%%%%%%%%
\begin{figure}[b]
 \begin{center}\vspace {0cm}
  \includegraphics*[angle=270,scale=0.32]{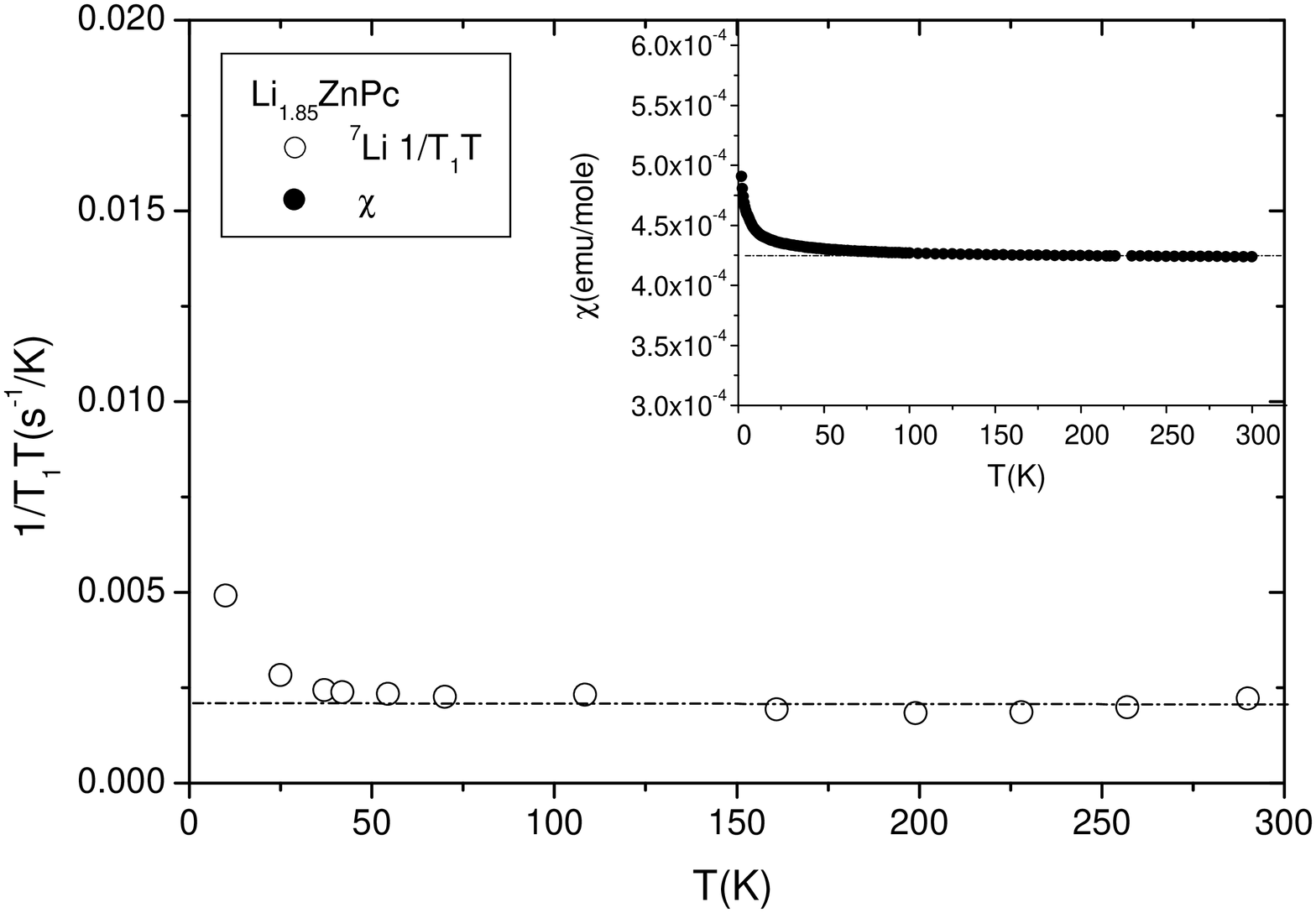}\caption{$^{7}$Li $1/T_1T$ in
Li$_{1.85}$ZnPc as functions of T. The dashed-dotted line is
  a guide to the eye. (Inset) T-dependence of the molar magnetic susceptibility of Li$_{1.85}$ZnPc
powders, measured with a SQUID magnetometer in a field of 1000 Gauss. The diamagnetic contributions have already
been subtracted from the raw data.}\label{1ovt1t}
\end{center}
\end{figure}
%%%%%%%%%%%%%%%%%%%%%%%%%%%%%%%%%%%%%%%%%%%%%%%%%%%%%%%%%%%%%%%%%%%%%%
Now, one can compare the $1/T_1T$ data of the different \lxpc
samples with $\chi_S$ data to evaluate the evolution of the
spectral weight of $\vec{q}=0$ and $\vec{q}= 2\vec{k}_F$
excitations on cooling. At high T $1/T_1T$, in all compounds, is
dominated by the first term of Eq. (\ref{beechcomplt1}) and the
curves tend to a constant non zero value proportional to
$\chi_S^2$ (inset of Fig. 2), decreasing with increasing $x$, for $x>1.85$. Remarkably,
in all \lxpc compounds the electrons are delocalized at RT. In
fact, if in Li$_{3.75}$ZnPc the electrons were localized, one
would expect $1/T_1T\rightarrow 0$ on increasing T, as $1/T_1$ is
constant in a paramagnetic insulator \cite{Boucher}. $1/T_1T$ in
Li$_{1.85}$ZnPc remains practically constant on cooling (Fig. 3)
with a trend similar to the one of $\chi$, which indicates that
the $\vec{q}=0$ excitations are dominant down to low T.
On the contrary, the growth observed in $1/T_1$ for Li$_{2.7}$ZnPc
and Li$_{3.75}$ZnPc has rather to be associated with the enhancement of the $\vec{q}=
2\vec{k}_F$ fluctuations leading to $\chi(2\vec{k}_F)\propto {({T}/{E_F})}^{K_{\rho}-1}\gg
\chi(q\simeq 0)$ \cite{Solyom, behnia}, with $K_{\rho}$ an exponent
characterizing both charge and spin correlation functions. In particular, since $1/T_1$ is constant for $T\geq 150$ K,  $K_{\rho}\rightarrow 0$ for $x\simeq 3.75$, i.e. $U/W\gg1$. The low-T
dynamics observed by $^{7}$Li NMR in Li$_{3.75}$ZnPc could then
originate from localized electrons, which eventually give rise to a
Spin Density Wave (SDW) as in the Bechgaard salts \cite{Jerome}.
%%%%%%%%%%%%%%%%%%%%%FIG5%%%%%%%%%%%%%%%%%%%%%%%%%%%%%%%%%%%%%%%%%%%%
\begin{figure}[b]
 \begin{center}\vspace {0cm}
  \includegraphics*[angle=270,scale=0.36, viewport=49 75 560 807]{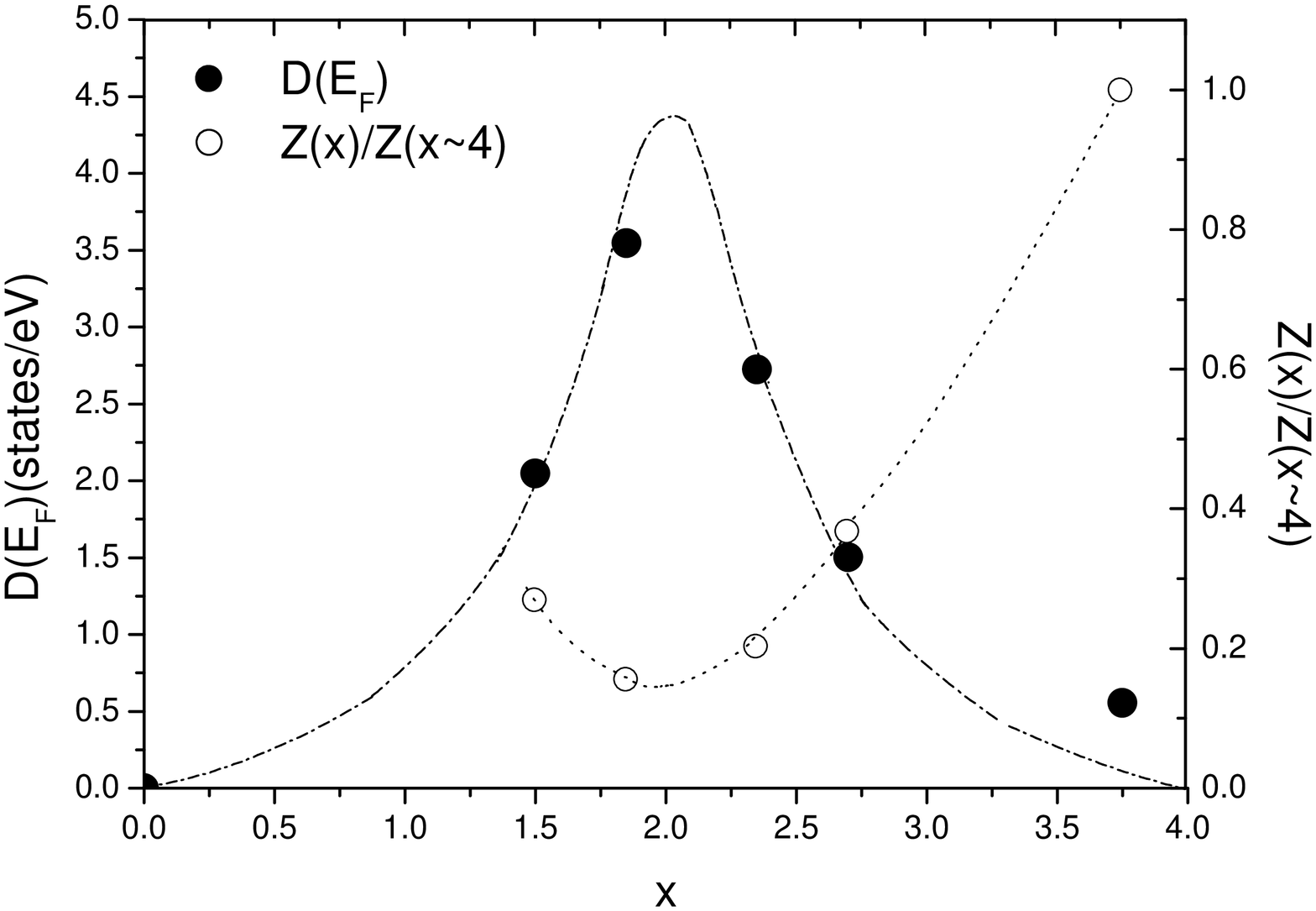}\caption{Density of states at the Fermi level
  per electron in \lxpc, obtained from susceptibility data, and the quasi-particle residue $Z(x)$,
  normalized to the value $Z(x\sim
4)$. The lines are guides to eye.}\label{dos}
\end{center}\end{figure}
%%%%%%%%%%%%%%%%%%%%%%%%%%%%%%%%%%%%%%%%%%%%%%%%%%%%%%%%%%%%%%%%%%%%%%

Now, the results obtained by means of the different techniques can be conveniently combined to derive the effective
 $D(E_F)$ as a function of Li content.
Since at high T the behaviour of $\chi_S$ is Pauli-like,
$\chi_{S}= N_A\mu_B^2 D(E_F)$, and since $1/T_1$ follows Eq.
(\ref{Korringaznpc}), one can quantitatively derive $D(E_F)$.
However, although all EPR measurements were performed in the same
experimental conditions, the absolute value of the EPR
susceptibility was not calibrated and, moreover, to derive
$D(E_F)$ from $1/T_1$ the knowledge of the hyperfine coupling $A$
is required. For $x= 1.85$ $A$ can be calculated by resorting to
Eq. (\ref{Korringaznpc}), on the basis of $1/T_1$ data and from
the susceptibility estimated with the SQUID magnetometer. One
finds $A(x=1.85)\simeq 2$ kGauss and $D(E_F)\simeq 3.6$ states/eV.
Upon increasing $x$ one finds that $C_0= 1/T_1\chi_S^2$ (see Eq.
1) slightly increases. The variation of this constant is related
to the variation of the hyperfine coupling $A$ with $x$. In
particular, one has $A(x=3.75)=1.32 A(x=2.7)= 1.27 A(x=1.85)$. Now
that the hyperfine constants are known, $D(E_F)$ can be derived
from $1/T_1$ or, alternatively, by scaling the absolute value of
$\chi_S$ determined by EPR with $\chi_{S}(x=1.85)= N_A\mu_B^2
D(E_F)$.

The doping dependence of $D(E_F)$ per electron derived from the
high T data is finally reported in Fig. \ref{dos}. It is noticed
that  $D(E_F)$ shows a marked peak for $x=2$, while it diminishes
almost symmetrically for $x\rightarrow 0$ and $x\rightarrow 4$.
Remarkably, $D(E_F)$ is finite for every $x\neq 0$ or 4,
indicating that electron injection in the $2e_g$ band always
drives the system towards metallicity. This is to our knowledge
the first clear indication that bulk MPc's become metallic upon
alkali doping despite of the significant Coulomb repulsion, in
agreement with the conductivity measurements in K$_x$MPc thin
films \cite{Morpurgo}. The increase by almost an order of
magnitude in the effective $D(E_F)$ at half-band filling cannot be
justified in the light of the bare $D(E)$. Recent band structure
calculations for K$_x$CuPc \cite{giovannetti} show an asymmetric
shape for the $D(E)$ in the $e_g$ bands. On the other hand, \lxpc
behave as a Fermi liquids with an effective bandwidth W$^*$ which
is sizeably reduced for $x\simeq 2$.  This narrowing was explained
by Capone \etal  within the Dynamical Mean Field Theory
\cite{Tosattiscience}. These authors have shown that in the strong
correlation limit, $0.8<U/W<0.9$, these systems are characterized
by an effective renormalized bandwidth $W^*(x)= Z(x)W$, with W
the bare bandwidth and Z the free fermion-like quasiparticle
residue which should vanish for U/W$\rightarrow 0.8$, i.e. close
to the MIT. The doping dependence of $Z$ derived from the one of
$D(E_F)$ is reported in Fig. \ref{dos} for comparison. One notices
that $Z$ is significantly reduced for $x\simeq 2$. Now, since
A$_x$MPc shows analogies with A$_x$C$_{60}$, one could envisage
the onset of strongly correlated superconductivity for $x\simeq
2$, with an effective coupling $J_{eff}= J-E_{JT}/4$. So far, for
$x\simeq 2$ no evidence of a superconducting transition  emerges
from the experimental data down to liquid helium T. Nevertheless,
it has to be remarked that this transition could be suppressed
either by a certain disorder or by the small inter-chain coupling
typical of these quasi-1D systems. Finally we point out that
recent infrared spectra show an increase in the reflectivity of
\lxpc powders upon Li doping, with a minimum around 0.6 eV for
$x\simeq 2.35$, possibly marking the plasma frequency.
\\In conclusion, from a series of NMR, EPR and magnetization
measurements we have evidenced that bulk \lxpc become metallic
upon Li doping. The T dependence of the nuclear
spin-lattice relaxation rate $1/T_1$ and of the static uniform
susceptibility $\chi_S$ on approaching room temperature are
characteristic of a Fermi liquid, with an effective $D(E_F)$ markedly peaked at half band
filling. While for $x\simeq 2$ the electrons are delocalized down
to low T, for $x\rightarrow 4$ a tendency towards
localization is noticed upon cooling. The $x$-dependence of
$D(E_F)$ shows that for $x\simeq 2$ \lxpc is close to a
metal-insulator transition where enhanced superconducting
fluctuations could develop, suggesting that in A$_x$MPc's a
phenomenology analogous to the one observed in A$_x$C$_{60}$ could
develop.
\\
\section*{Acknowledgements}
We are particularly grateful to M. Galli for letting us report on
his recent results with IR spectroscopy. Enlightening discussions
M. Fabrizio and E. Tosatti are also acknowledged. This work was
supported by Fondazione CARIPLO 2005 Scientific Research funds.

%%%%%%%%%%%%%%%%%%%%%%%%%%%%%

%%%%%%%%%%%%%%%%%%%%%%%%%%%%%%%%%%%%

\end{document}